\documentclass{aa}
\usepackage{graphicx}
\usepackage{txfonts}
\def\arcmin{\hbox{$^\prime$}}
\def\arcsec{\hbox{$^{\prime\prime}$}}

\begin{document}
   \title{Optical observations of GRB 060124 afterglow: A case for an injection break}

   \author{Kuntal Misra
          \inst{1},
	  D. Bhattacharya
	  \inst{2},
	   D. K. Sahu
	  \inst{3},
	   Ram Sagar
	  \inst{1},
	   G. C. Anupama
	  \inst{4},
	  A. J. Castro-Tirado
          \inst{5},
          S. S. Guziy
          \inst{5,6}
          \and
	  B. C. Bhatt
	  \inst{3}
          }
          %\fnmsep\thanks{Just to show the usage
          %of the elements in the author field}

   \offprints{Kuntal Misra}

   \institute{Aryabhatta Research Institute of Observational Sciences,
              Manora Peak, Nainital 263 129, India\\
              \email{kuntal,sagar@aries.ernet.in}
               \and
             Raman Research Institute, Bangalore 560 080, India\\
             \email{dipankar@rri.res.in}
		\and
		Center for Research and Education in Science \& Technology, Hosakote, Bangalore 562 114, India\\
              \email{dks,bcb@crest.ernet.in}
               \and
		Indian Institute of Astrophysics, Bangalore, 560 034, India\\
              \email{gca@iiap.res.in}
		\and
               Instituto de Astrof\'{\i}sica de Andaluc\'{\i}a (IAA-CSIC), PO Box 03004, 18080 Granada, Spain\\
               \email{ajct,gss@iaa.es}
		\and
                Nikolaev State University, Nikolskaya 24, 54030 Nikolaev, Ukraine\\
             }

   \date{Received ; accepted }

% \abstract{}{}{}{}{}
% 5 {} token are mandatory
 \abstract
  % context heading (optional)
  % {} leave it empty if necessary
{}
  % aims heading (mandatory)
{We present broad band optical afterglow observations of a long duration
GRB 060124 using the 1.04-m Sampurnanand Telescope at ARIES, Nainital and
the 2.01-m HCT at IAO, Hanle, including the earliest ground based observations
in R band for this GRB. We determine the decay slope of the light curve at
different bands and examine the reality of a proposed jet break.}
  % methods heading (mandatory)
{We use data from our observations as well as others reported in the
literature to construct light curves in different bands and make power law
fits to them. The spectral slope of the afterglow emission in the optical
band is estimated.}
  % results heading (mandatory)
{Our first R-band observations were taken $\sim 0.038$~d after burst.
We find that all available optical data after this epoch are well fit by a
single power law, with a temporal flux decay index $\alpha\sim 0.94$.
We do not find any evidence of a jet break within our data, which extend
till $\sim 2$~d after the burst. The X-ray light curve, however, shows a
distinct break around 0.6 day. We attribute this break to a steepening
of the electron energy spectrum at high energies.}
  % conclusions heading (optional), leave it empty if necessary
{We conclude that the above measurements are consistent with the picture
of a standard fireball evolution with no jet break within $t\sim 2$~days
after the burst. This sets a lower limit of $3\times 10^{50}$~erg to the
total energy released in the explosion.}
  {}

   \keywords{gamma ray bursts --
                photometry --
                observations -- temporal index -- spectral index               }

   \authorrunning {Kuntal Misra et al.}
   \titlerunning {Optical Observations of GRB 060124 Afterglow}
   \maketitle
%
%________________________________________________________________

%--------------------------------------------INTRODUCTION------------------------------------------%
\section{Introduction}
GRB 060124 ({\it Swift} trigger = 178750), a long duration gamma ray burst, was detected on
2006 January 24 at T = 15:54:52 UT by the {\it Swift}-BAT (Holland et al. 2006a). The BAT light curve shows
a precursor from T-3 to T+13 sec, followed by three major peaks from T+520 to T+550 sec,
T+560 to T+580 sec, which has the largest flux, and T+690 to T+710 sec (Fenimore et al. 2006).
The total duration of the GRB is, thus, the longest ever recorded by {\it BATSE} or {\it Swift}.
The fluence of the precursor emission in the 15-150 keV band is (4.6 $\pm$ 0.5) x 10$^{-7}$ erg/cm$^2$.
The peak flux in the 15-150 keV band was about 4.5 $\pm$ 0.5 counts/cm$^2$/sec at T+570.
The estimated total fluence in the BAT energy range is $\sim$ 7 x 10$^{-6}$ erg/cm$^2$ in the 15-150 keV
band after scaling the fluence of the precursor.

{\it Swift}-XRT began observing the field 106 sec after the BAT trigger. Analysis of the first orbit of
XRT data shows the presence of an X-ray source at a position
$\alpha_{2000}$ = $05^{\rm h} 08^{\rm m} 27^{\rm s}.27$, $\delta_{2000}$ = $+69^{\circ} 44{\arcmin} 25{\arcsec}.7$
which is 2.4 arcmin from the BAT position and 9 arcsec from the optical afterglow candidate reported by
Kann (2006). The X-ray light curve in the Windowed Timing mode shows an initial flat
behavior followed by three bright flares after the first 100 s of observation (Mangano et al. 2006).

An optical afterglow candidate of the GRB 060124 was discovered by Kann (2006) at
$\alpha_{2000}$ = $05^{\rm h} 08^{\rm m} 25^{\rm s}.5$, $\delta_{2000}$ = $+69^{\circ} 44{\arcmin} 26{\arcsec}$.
V-band exposures taken 184 and 629 sec after the BAT trigger by {\it Swift}-UVOT also show the presence of
an optical transient (Holland et al. 2006b). No optical flare was detected at the position of the
optical afterglow (Torii 2006).
A low-resolution spectrum of the afterglow of GRB 060124 was obtained by Mirabal and Halpern (2006) on
2006 Jan 25.13. Its preliminary analysis shows the presence of a significant absorption feature at
5105 $\AA$, possibly a doublet corresponding to Mg II 2795 $\AA$, 2802 $\AA$ (z=0.82) or C IV 1548 $\AA$,
1550 $\AA$ (z=2.30).
The presence of doublet feature is later confirmed by Cenko et al. (2006). However, the Mg II
feature has been ruled out on the basis of their separation and is identified as CIV feature by
Prochaska et al. (2006). They also identified a weak absorption feature consistent with this redshift corresponding
to AlII 1670 $\AA$ and an absorption line at ~4000$\AA$ consistent with Ly$\alpha$ absorption and indicating
log N(HI) $<$ 20.5.
According to Prochaska et al. (2006) the afterglow spectrum is notable for exhibiting very weak low-ion features
(e.g. non-detections of FeII 1608, OI 1302, CII 1334) and relatively low HI column density. In this respect, the
spectrum of GRB 060124 afterglow is similar to the afterglows of GRB 050908 and GRB 021004.
Assuming a standard cosmological model with $H_{0}$ = 70 km/s/Mpc, $\Omega_{\rm M}$ = 0.3, $\Omega_{\Lambda}$ =
0.7 and a redshift of z=2.297 and a fluence of about 7 x 10$^{-6}$ erg/cm$^2$, the isotropic-equivalent gamma
ray energy is 8.9 x 10$^{52}$ erg (Cenko et al. 2006).

The most intense peak of the burst as seen by {\it Swift} BAT triggered {\it Konus-Wind} 558.9 s
after the BAT trigger (Golenetskii et al. 2006).
The {\it Konus-Wind} light curve shows the presence of the precursors and the three major peaks as were seen by
{\it Swift} BAT. Golenetskii et al. (2006) make a preliminary estimate of the total burst fluence in the
20 keV - 2 MeV energy range to be $\sim$ 2.80 $\times$ 10$^{-5}$ erg/cm$^{2}$/sec.
The major peak of GRB 060124 was intense enough to trigger the FREGATE instrument on {\it HETE II}
557.7 s after the {\it Swift} trigger time (Lamb et al. 2006).

GRB 060124 has been the first well studied event by {\it Swift} in the prompt and the afterglow emission phase
(Romano et al. 2006). The prompt emission was simultaneously observed by XRT and UVOT. GRB 990123 was one
such event for which the flare in the prompt phase was observed by ROTSE (Akerlof et al. 1999) which showed
a rapid decline. This emission in the prompt phase was interpreted as reverse shock emission.
The {\it Konus-Wind} light curve resembles the light curves of two long duration bursts:
GRB 041219A (Vestrand et al. 2005) and GRB 050820A (Golenetskii et al. 2005).
The IR afterglow of GRB 041219A was observed while the GRB was still going on.
Optical imaging of GRB 050820A, 5.5 s after the trigger alert, with the RAPTOR telescope clearly shows the
emergence of faint optical emission which suddenly flares and fades away during the first one hour
(Vestrand et al. 2006).
The prompt optical emission in both these long duration GRBs (GRB 041219A and GRB 050820A)
has been attributed to internal shocks in the ultra-relativistic outflow
(Vestrand et al. 2005, Vestrand et al. 2006), the same source that is
thought to generate the gamma ray emission in the burst.
The optical afterglow of GRB 060124 was extensively followed by ground based telescopes with the
earliest observations reported by our group (Misra 2006).

In this paper we present the broad band $BVRI$ photometric observations of the optical afterglow of GRB 060124
including the earliest observations in R band from any ground based telescope. A secure photometric calibration
has been carried out by imaging the Landolt (1992) Standard SA 104 field along with the GRB 060124 field.
A brief description about the observations and data reduction is presented in the next section whereas
the multi-band optical light curve of the afterglow follow in section 3.
The discussions and conclusions form the last section of the paper.

%------------------------------OBSERVATIONS AND DATA REDUCTION-----------------------------------%
\section{Observations and Data Reduction}
Optical observations of the GRB 060124 afterglow were carried out using the 1.04-m Sampurnanand Telescope (ST)
at Aryabhatta Research Institute of observational sciencES (ARIES), Nainital
and the 2.01-m Himalayan Chandra Telescope (HCT) at Indian Astronomical Observatory (IAO), Hanle during
2006 January 24 to 26.
A CCD chip of size 2048 $\times$ 2048 pixel$^2$ was used at ST covering a field of
$\sim$ 13$\arcmin$ $\times$ 13$\arcmin$ on the sky.
The gain and readout noise of the CCD camera are 10 e$^-$/ADU and 5.3 electrons respectively.
The filters used at ST are the Johnsons BV and Cousins RI.
The frames were binned in 2 $\times$ 2 pixel$^2$ to improve the signal-to-noise ratio of the source.
The CCD used at HCT was a 2048 $\times$ 4096 SITe chip mounted on Himalayan Faint Object Spectrograph
Camera (HFOSC). The central 2048 $\times$ 2048 region of the CCD was used for imaging
which covers a field of 10$\arcmin$ $\times$ 10$\arcmin$ on the sky.
It has a gain of 1.22 e$^-$/ADU and readout noise of 4.8 electrons. Filters used are Bessells R and V.
The frames obtained using HCT were unbinned.
Several twilight flat field and bias frames were obtained for the CCD images at both the telescopes.
Several short exposures, with exposure time varying from 300 sec to 1800 sec, were taken in different
filters to image the optical transient (OT) of GRB 060124.

MIDAS, IRAF and DAOPHOT softwares were used to process the CCD frames in standard fashion.
The bias subtracted, flat fielded and cosmic ray removed CCD frames were co-added whenever
found necessary.

\begin{figure}
   \centering
\includegraphics[width=8cm,height=8cm]{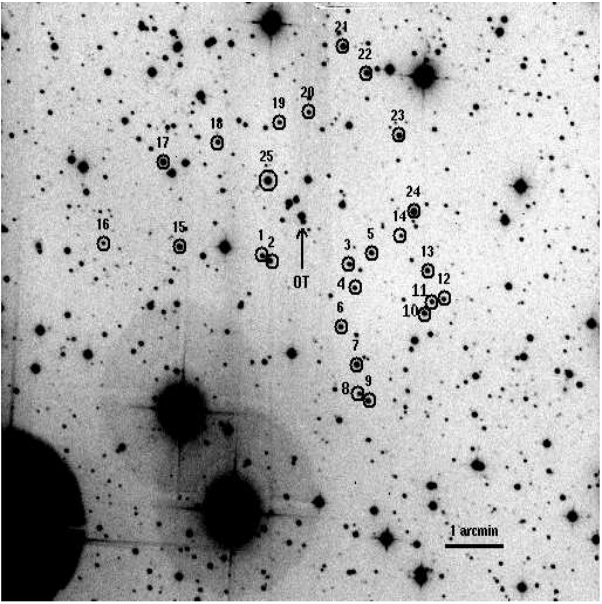}
      \caption{A R-band image of GRB 060124 optical afterglow with the 1.04-m Sampurnanand Telescope (ST) at
ARIES, Nainital. The optical afterglow is indicated by an arrow. Marked are the 25 secondary stars used
for calibration. North is up and East is to the left. The approximate scale is shown towards the bottom
right corner.
              }
         \label{st_aries_060124}
   \end{figure}

Landolt (1992) Standard field SA 104 region was imaged to calibrate the field of GRB 060124.
Atmospheric extinction values and the transformation coefficients in B, V, R and I filters
were determined using the 6 standard bright stars in the SA 104 region.
The 6 standard stars in the SA 104 region cover a range
of 0.518 $< (B-V) <$ 1.356 in color and 13.484 $< V <$ 16.059 in brightness.
Using these transformation, $BVRI$ magnitudes of 25 secondary stars
were determined in GRB 060124 field and their average values are listed in Table 1. Figure \ref{st_aries_060124}
shows the
position of the optical afterglow in R band and the 25 secondary stars used for calibration.
These stars have internal photometric accuracy better than 0.01 mag.
The (X, Y) CCD pixel coordinates were converted to $\alpha_{2000}$, $\delta_{2000}$ values using the
astrometric positions given by Henden (2006). A comparison between photometry of Henden (2006) and that of
ours yields zero point differences of 0.04 $\pm$ 0.03, 0.04 $\pm$ 0.03, 0.03 $\pm$ 0.02 and 0.04 $\pm$ 0.03
in $B$, $V$, $R$, $I$ filters respectively. These zero point differences are based on the comparison of
14 common secondary stars in the field of GRB 060124.

\begin{figure}
   \centering
\includegraphics[width=8cm,height=8cm]{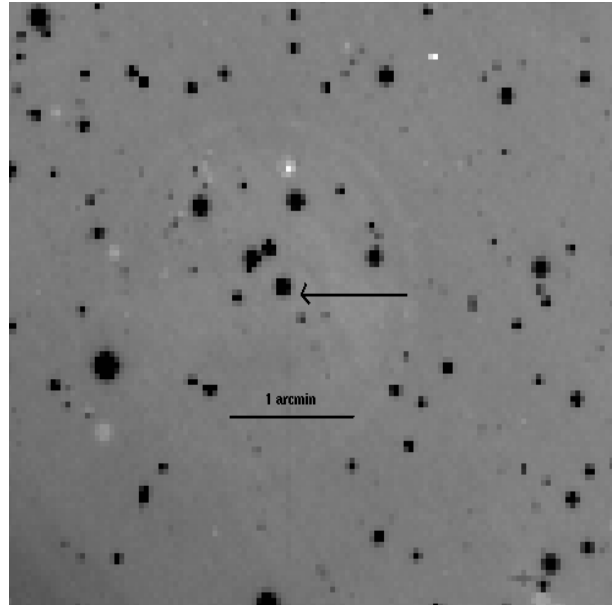}
      \caption{A R-band image of the field of GRB 060124 taken on 2006 February 16 (22 days after the burst)
with the 2.5-m Issac Newton Telescope (INT) at La Palma, Spain which does not show anything at the position of the
optical afterglow. North is up and East is to the left. The approximate scale is shown in the figure.
              }
         \label{int_060124}
   \end{figure}

The observations of the field of GRB 060124 were carried out towards a later epoch on 2006 February 16
with the 2.5-m Issac Newton Telescope (INT) at La Palma, Spain. Figure \ref{int_060124} shows the zoomed area of
the field of GRB 060124 which does not reveal any afterglow candidate at the location of the optical
transient.

A full compilation of $BVRI$ magnitudes of the optical afterglow differentially calibrated with respect
to the secondary stars numbered 1, 2, 3, 4, 5 and 6 as listed in Table 1 is presented in Table 2.
The photometric magnitudes reported by Masetti et al. (2006) were converted to the present photometric
scale using the secondary stars listed in Table 1.

%---------------------------------Figure 1---------------------------------%
%______________________________________________________________

%---------------------------------Figure 2---------------------------------%
%______________________________________________________________

%------------------------------TABLE 1: Calibration Stars-----------------------%
\begin{table}
\caption{The identification number (ID), ($\alpha$, $\delta$) for epoch 2000, standard $V$, $(B-V)$, $(V-R)$ and $(R-I)$ photometric magnitudes of the secondary standards in the GRB 060124 field}
\label{table:1}      % is used to refer this table in the text
\centering                          % used for centering table
\begin{tabular}{c c c c c c c}        % centered columns (7 columns)
\hline\hline                 % inserts double horizontal lines
ID & $\alpha_{2000}$ & $\delta_{2000}$ & $V$ & $(B-V)$ & $(V-R)$ & $(R-I)$\\
& (h m s)&(deg m s)&(mag)&(mag)&(mag)&(mag)\\   % table heading
\hline                        % inserts single horizontal line
1 & 05 07 41 &69 41 02 & 16.99 & 1.41 & 0.92 & 0.83\\
2 & 05 07 59 &69 41 48 & 18.83 & 1.61 & 1.19 & 1.42\\
3 & 05 07 57 &69 42 21 & 15.45 & 0.99 & 0.53 & 0.49\\
4 & 05 07 54 &69 42 52 & 17.05 & 1.08 & 0.63 & 0.59\\
5 & 05 08 08 &69 41 12 & 17.06 & 0.82 & 0.45 & 0.45\\
6 & 05 08 11 &69 41 51 & 16.66 & 0.86 & 0.49 & 0.47\\
7 & 05 08 15 &69 42 32 & 17.12 & 0.74 & 0.39 & 0.41\\
8 & 05 08 12 &69 43 16 & 18.08 & 0.82 & 0.45 & 0.44\\
9 & 05 08 33 &69 43 42 & 16.78 & 0.95 & 0.51 & 0.46\\
10& 05 08 27 &69 44 51 & 16.83 & 1.42 & 0.89 & 0.77\\
11& 05 08 25 &69 45 14 & 15.27 & 0.77 & 0.42 & 0.41\\
12& 05 08 35 &69 45 10 & 14.93 & 0.72 & 0.39 & 0.39\\
13& 05 08 31 &69 44 28 & 18.09 & 0.95 & 0.53 & 0.46\\
14& 05 08 50 &69 43 43 & 16.56 & 0.88 & 0.49 & 0.46\\
15& 05 08 56 &69 43 52 & 16.34 & 0.86 & 0.47 & 0.45\\
16& 05 09 15 &69 43 51 & 19.18 & 1.49 & 1.04 & 1.07\\
17& 05 09 01 &69 45 25 & 15.71 & 0.69 & 0.38 & 0.39\\
18& 05 08 47 &69 45 50 & 17.41 & 0.67 & 0.37 & 0.39\\
19& 05 08 32 &69 46 16 & 17.89 & 0.75 & 0.43 & 0.43\\
20& 05 08 25 &69 46 29 & 18.18 & 0.85 & 0.48 & 0.49\\
21& 05 08 17 &69 47 44 & 16.09 & 0.93 & 0.51 & 0.48\\
22& 05 08 11 &69 47 16 & 16.01 & 0.67 & 0.38 & 0.38\\
23& 05 08 02 &69 46 08 & 16.64 & 1.07 & 0.58 & 0.57\\
24& 05 07 58 &69 44 44 & 15.40 & 0.63 & 0.35 & 0.38\\
25& 05 07 54 &69 43 40 & 17.18 & 0.99 & 0.56 & 0.53\\
\hline                                   %inserts single line
\end{tabular}
\end{table}
%______________________________________________________________

%-----------------------------------------------TABLE 2: Observation Log----------------------%
\begin{table}
\caption{Log of GRB 060124 Afterglow observations}
\label{table:2}      % is used to refer this table in the text
\centering                          % used for centering table
\begin{tabular}{c c c c c}        % centered columns (7 columns)
\hline\hline
Date (UT) & Time since &Magnitude & Exposure& Tel.\\
2006 January&burst (days)& (mag)&Time (s)&\\   % table heading
\hline                        % inserts single horizontal line
&\multicolumn{3}{c}{\bf{$B$-passband}}\\
25.6434&  0.9802 & 20.58 $\pm$ 0.048 & 1200 & ST\\
25.6925&  1.0294 & 20.69 $\pm$ 0.042 & 1200 & ST\\
25.7279&  1.0648 & 20.67 $\pm$ 0.051 & 1200 & ST\\
25.7624&  1.0993 & 20.78 $\pm$ 0.043 & 1200 & ST\\
26.6542&  1.9911 & 21.30 $\pm$ 0.075 & 1800 & ST\\
26.7194&  2.0563 & 21.54 $\pm$ 0.070 & 1800 & ST\\
&\multicolumn{3}{c}{\bf{$V$-passband}}\\
25.6314&  0.9683 & 20.14 $\pm$ 0.037 & 600  & ST\\
25.6372&  0.9741 & 20.15 $\pm$ 0.029 & 600  & HCT\\
25.6420&  0.9789 & 20.13 $\pm$ 0.032 & 600  & HCT\\
25.6653&  1.0022 & 20.19 $\pm$ 0.045 & 600  & ST\\
25.6771&  1.0140 & 20.20 $\pm$ 0.036 & 1200 & ST\\
25.7161&  1.0530 & 20.25 $\pm$ 0.057 & 600  & ST\\
25.7506&  1.0875 & 20.28 $\pm$ 0.053 & 600  & ST\\
25.7830&  1.1199 & 20.36 $\pm$ 0.032 & 600  & HCT\\
26.6190&  1.9559 & 20.96 $\pm$ 0.052 & 1200 & ST\\
26.6306&  1.9675 & 20.89 $\pm$ 0.038 & 3x600& HCT\\
26.6729&  2.0098 & 21.17 $\pm$ 0.071 & 1200 & ST\\
26.7418&  2.0787 & 21.00 $\pm$ 0.064 & 1800 & ST\\
26.8806&  2.2175 & 20.89 $\pm$ 0.048 & 3x600& HCT\\
&\multicolumn{3}{c}{\bf{$R$-passband}}\\
24.7017 & 0.0386 & 16.83 $\pm$ 0.007 & 300  & ST\\
24.7069 & 0.0438 & 16.93 $\pm$ 0.006 & 300  & ST\\
24.7114 & 0.0483 & 17.02 $\pm$ 0.008 & 300  & ST\\
24.7201 & 0.0570 & 17.23 $\pm$ 0.015 & 600  & HCT\\
24.7288 & 0.0657 & 17.37 $\pm$ 0.011 & 600  & HCT\\
24.7373 & 0.0742 & 17.48 $\pm$ 0.010 & 600  & HCT\\
25.5594 & 0.8963 & 19.80 $\pm$ 0.029 &2x600 & HCT\\
25.5820 & 0.9189 & 19.61 $\pm$ 0.055 & 300  & ST\\
25.5871 & 0.9240 & 19.77 $\pm$ 0.042 & 300  & ST\\
25.6083 & 0.9452 & 19.67 $\pm$ 0.050 &2x600 & HCT\\
25.6113 & 0.9482 & 19.83 $\pm$ 0.055 & 300  & ST\\
25.6264 & 0.9633 & 19.86 $\pm$ 0.020 & 600  & HCT\\
25.6581 & 0.9950 & 19.93 $\pm$ 0.065 & 300  & ST\\
25.7089 & 1.0458 & 19.88 $\pm$ 0.056 & 400  & ST\\
25.7432 & 1.0801 & 19.84 $\pm$ 0.071 & 400  & ST\\
25.7576 & 1.0945 & 19.85 $\pm$ 0.033 & 600  & HCT\\
25.7660 & 1.1029 & 19.97 $\pm$ 0.025 & 600  & HCT\\
26.5615 & 1.8984 & 20.58 $\pm$ 0.030 &3x600 & HCT\\
26.6308 & 1.9677 & 20.66 $\pm$ 0.071 & 600  & ST\\
26.7647 & 2.1016 & 20.55 $\pm$ 0.047 & 1800 & ST\\
Feb 16.5146& 22.9086& 22.50             & 3x600 & INT\\
&\multicolumn{3}{c}{\bf{$I$-passband}}\\
25.5918 & 0.9287 & 19.52 $\pm$ 0.122 & 300 & ST\\
25.6002 & 0.9371 & 19.20 $\pm$ 0.061 & 300 & ST\\
25.6049 & 0.9418 & 19.07 $\pm$ 0.068 & 300 & ST\\
25.6533 & 0.9902 & 19.25 $\pm$ 0.071 & 300 & ST\\
25.7030 & 1.0399 & 19.27 $\pm$ 0.079 & 400 & ST\\
25.7379 & 1.0748 & 19.67 $\pm$ 0.112 & 300 & ST\\
25.7729 & 1.1098 & 19.49 $\pm$ 0.145 & 400 & ST\\
26.6391 & 1.9760 & 20.19 $\pm$ 0.158 & 600 & ST\\
26.7000 & 2.0369 & 20.56 $\pm$ 0.208 & 1200& ST\\
\hline                                   %inserts single line
\end{tabular}
\end{table}
%______________________________________________________________

%--------------------------------------------RESULTS------------------------------%
\section{Multiband Optical Light Curve}
Figure 3 shows the temporal evolution of the GRB 060124 afterglow in $BVRI$ bands
based on the data presented here and those available in the literature.
The published magnitudes in the optical bands were brought to
our photometry level using the 6 secondary standard stars as mentioned in Section 2. The frequency distribution
of our data in $BVRI$ bands is $N(B, V, R, I)$ = (6, 13, 20, 9). It is to be noted here that
we present the earliest ground based R band observations which have placed important constraints on the
overall evolution of the afterglow. The X-axis here represents log $\Delta t (=t-t_{0})$ where $t$ is the
time of observation and $t_{0}$ = 2006 January 24.6331 UT is the burst epoch.

Since we have the R band observations from $\Delta t$ = 0.03 to 2.0 day, we noticed that
the R band flux decay of the afterglow of GRB 060124 can be well characterized by a single power law written as\\

\begin{figure}
   \centering
\includegraphics[width=8cm,height=8cm]{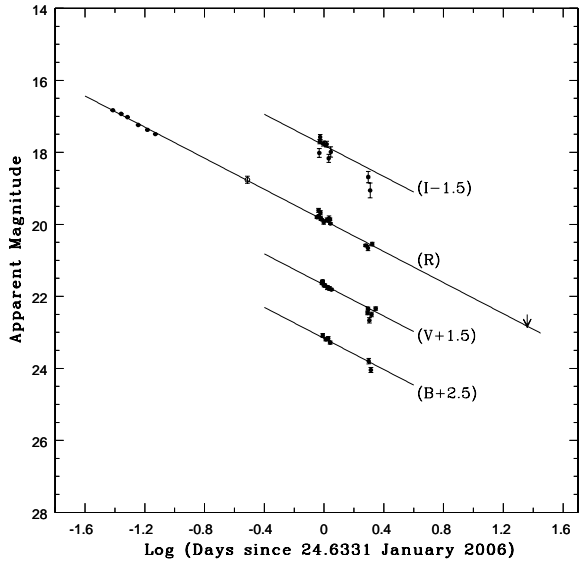}
      \caption{Optical light curve of the GRB 060124 afterglow in $BVRI$ bands. The light curves in different
bands are offset by a number as indicated to avoid overlap of the data. Filled circles represent the data from
the present measurements whereas unfilled circles represent the data taken from the literature. The solid
line indicates the least square best fitted value of the temporal index. Also shown is the upper limit derived
using the observations at a later epoch.
              }
         \label{optical_lc}
   \end{figure}

$F(t)$ $\propto$ $t^{-\alpha}$\\

\noindent
where $F(t)$ is the flux of the optical afterglow at time $t$ since the burst and $\alpha$ is the temporal
flux decay index.
The above function was fitted to the R band data using the least square regression method. This yields
the value of $\alpha$ = 0.94 $\pm$ 0.01. We assumed the same power law decay for B, V and I bands and fitted the
above function which seems to agree with the observed data points. The linear least square fit to the
data in the different bands is shown in figure \ref{optical_lc}.

%----------------------------------Figure 3-------------------------%
%--------------------------------------------------------------------

Our observations extend upto $\Delta t$ = 2 days after the burst including the earliest R band data at
$\Delta t$ = 0.03 day and it does not show any
noticeable steepening in the light curve  during this entire period of observations.

Towards a later epoch on 2006 February 16, the field of GRB 060124 was observed with the 2.5-m INT at
La Palma, Spain. We do not see the optical afterglow in a co-added frame of 3 $\times$ 600 sec
exposure at the location of the optical transient and thus put an upper limit of 22.5 mag at the
location of the afterglow.
There is also no sign of any galaxy at the location. The disappearance of the afterglow is apparent in
figure \ref{int_060124} when compared to figure \ref{st_aries_060124}.

Both optical and X-ray light curves of GRB 060124 obtained from {\it Swift} show a remarkable flickering
which can also be seen at late times in our optical observations at all the passbands.
Achromatic flickering in optical bands has been seen in a few other GRB
afterglows such as GRB 000301C (Sagar et al. 2000), GRB 021004 (Pandey et al.
2003, Bersier et al. 2003, Bj\"{o}rnsson et al. 2004), GRB 021211 (Li et al. 2003), GRB 050319
(Wozniak et al. 2005) and GRB 051028 (Castro-Tirado et al. 2006). While some of the reported flicker
could arise from
variable observing conditions (Wozniak et al 2005), most of the observed
flickering appears to be real. Short timescale variability could, in many
cases, be attributed to density variations in the external medium
(Wang \& Loeb 2000). However, in some cases late injection of energy
from the central engine provide a better explanation (e.g.\ Zhang and
Meszaros 2002, Li et al.\ 2003).
Gravitational microlensing has been proposed as the cause of achromatic
variability in GRB~000301C (Garnavich et al.\ 2000), but this interpretation
remains ambiguous.
In the case of GRB~060124, the flickering is observed in all the optical
bands as well as in X-rays. Time scales of this variability are short compared
to the elapsed time since burst, and there is no evidence seen of persistent
increase in the light curve level following the flicker episodes.  This nature
is more akin to the variability expected due to inhomogeneities in the
density of the circumburst material.

%------------------------------------DISCUSSIONS AND CONCLUSIONS------------------%
\section{Discussions and Conclusions}
We present in this paper the broad-band $BVRI$ optical observations using the 1.04-m Sampurnanand Telescope at ARIES,
Nainital and the 2.01-m HCT at IAO, Hanle during 2006 January 24 to 26. We report the earliest optical observations
in the R band at $\Delta t <$ 1 hr. The X-ray light curve showed a ``flaring'' prompt emission. The prompt
emission was also observed at optical wavelengths but no flaring activity was seen.
The available data in the optical band ({\it Swift} UVOT + ground based observations) is
not sufficient to compare the prompt optical emission with that in gamma ray and X-ray bands.

We study the optical afterglow evolution from 0.03 to 2.0 day. The optical afterglow is well characterized
with a single power law decay with a flux decay index of 0.94 $\pm$ 0.01.
The late INT upper limit shows that any contamination from possible steady sources coincident with the afterglow
is fainter by at least 2 magnitudes compared to the last observed afterglow magnitude at $\sim$ 2 days. This ensures
that steepening of the optical light curve, if any, within 2 days could not be masked by a steady source
contribution.
Table 3 lists the colors of the afterglow at $\Delta t \sim$ 1 and 2 days after the burst.
The spectral index ($\beta_{\rm opt}$) of the afterglow of GRB 060124 in the optical band
is estimated using the flux normalization at different frequencies from the light curves.
The estimated value of Galactic Interstellar Extinction in the direction of the burst, using the reddening
map provided by Schlegel et al. (1998) is E(B-V) = 0.135 mag. The apparent magnitudes, corrected for Galactic
extinction only as the extinction due to the host galaxy is unknown, were converted to flux using the
normalization in the $BVRI$ bands by Bessel et al. (1998).
The spectrum is described by a single power law $F_{\nu}$ $\propto$ $\nu^{-\beta}$, where $F_{\nu}$ is the flux at a
frequency $\nu$ and $\beta$ is the spectral index. The value of $\beta_{\rm opt}$ obtained in this way is 0.73 $\pm$ 0.08.

%----------------------------------------TABLE: 3----------------------------------%
\begin{table}
\caption{Colors of the GRB 060124 Afterglow at $\delta t$ = 1.04 and 1.99 day}
\label{table:3}      % is used to refer this table in the text
\centering                          % used for centering table
\begin{tabular}{c c c c c}        % centered columns (7 columns)
\hline\hline                 % inserts double horizontal lines
$\delta t$ & $(B-R)$  & $(B-V)$ & $(V-R)$ &  $(R-I)$\\
days&mag&mag&mag&mag \\
\hline   % table heading
1.04&0.79$\pm$0.05 & 0.41$\pm$0.05 & 0.37$\pm$0.06 & 0.61$\pm$0.06 \\
1.99&0.64$\pm$0.08 & 0.13$\pm$0.08 & 0.51$\pm$0.07 & 0.47$\pm$0.09 \\
\hline                        % inserts single horizontal line
\end{tabular}
\end{table}
%______________________________________________________________

Romano et al. (2006) tried to fit the XRT light curve beyond 10$^4$ sec with a single power law which gives
a flux decay index of $\alpha$ = 1.36 $\pm$ 0.02. But they find a significant improvement
in the fit with a broken power law with a break time $t_b$ = 1.05$^{+0.17}_{-0.14}$
$\times$ 10$^5$ sec, $\alpha_{1}$ = $1.21 \pm 0.04$ and $\alpha_{2}$ = $1.58 \pm 0.06$ ($\Delta \alpha$ = 0.37).
Romano et al. (2006) consider this break in the XRT light curve too shallow
to be interpreted as a jet break. The optical UVOT data from {\it Swift} is available only for $t < 10^{5}$ s and
hence cannot be used to examine a possible break around this epoch.

Curran et al. (2006) also mention a break in the X-ray and optical light curve at 0.77 day but they do not quote the
quality of the fit. They also mention that the achromatic break seen in the X-ray and optical light curves
differ significantly from the usual jet break which is apparent in GRB afterglows. The late time temporal decay
rates reported by Curran et al. (2006) are unequal in X-ray and optical bands, and both are shallower than
expected in the case of a regular jet break.

With our coverage of optical data, we also try to check for a possible jet break by fitting a double power law
to the observed data.
However, combining the early and late time observations in R band we do not find any convincing
evidence for a jet break around $10^{5}$s as seen in the X-ray light curve by Romano et al. (2006) and
there is no considerable
steepening seen in the light curve till $\sim$ 2 days after the burst. Using the same values of $\alpha_1$,
$\alpha_2$ and $t_b$ as in the X-ray band (Romano et al. 2006), we attempt a broken power law fit to the
R band light curve. This yields a very high value of $\chi^2$ / d.o.f. = 269 which clearly rules out
the possibility of a similar break in the optical light curve.
Even the inclusion of a possible contribution from a steady source, consistent with the INT upper limit,
does not improve the broken power law fit appreciably.
Similarly using the values reported by
Curran et al. (2006) for the optical data, of break time $t_b$ = 0.77 day, $\alpha_1$ = 0.77 and $\alpha_2$ = 1.30
we fit a broken power law to the R band light curve. This fit is ruled out at a high confidence level with a
$\chi^2$ / d.o.f. = 15.
Thus, we conclude that the optical light curve is well characterized by a single power law.

Based on the fits made to the XRT light curve by Romano et al. (2006) and optical light curve (present work),
we see a break which is present in the XRT light curve but not in the optical light curve. Thus, this is
a frequency dependent break and is therefore not of dynamical origin. Steepening
of this kind reflects a break in the energy distribution of the radiating electrons. Such a break,
most commonly, is attributed to the effect of synchrotron cooling. Synchrotron cooling results in the
steepening of the slope of the electron energy distribution by one and a corresponding change in the
light curve slope by
0.25 (before the jet break). The break that is seen in the XRT light curve is however much steeper
than 0.25. We tried to fit the XRT light curve by fixing a light curve slope of 0.25
(similar to that required by a cooling break) and rule out such a fit with a high confidence level (best fit
$\chi^2$ / d.o.f. = 31). The steepening observed in the XRT light curve must therefore be attributed to a
different cause. It might be that the energy spectrum of the injected electrons steepens beyond a certain high
energy limit (see, eg. Panaitescu and Kumar 2001). Such an ``injection break" as it passes through an
observing frequency, leads to a corresponding break in the radiation spectrum, and to
a steepening of the light curve. We may model such an evolution using a general form of the expression
used by Wei and Lu (2002).
Wei and Lu (2002) propose that
the multi-wavelength spectra of GRB afterglows may not be made of exact power law segments but with a smooth
change in slope and
some breaks in the GRB afterglow light curves may be caused by such curved spectra.

For an injection break ($\gamma_i$) in the electron energy spectrum  evolving as $\gamma_i \propto \Gamma^q$,
where $\Gamma$ is the bulk Lorentz factor of the shock (see Bhattacharya 2001), we can describe the flux
evolution at a given observing frequency $\nu_{obs}$ as

\begin{equation}
F_{\nu_{\rm obs}}\propto
\frac{(\frac{t}{t_{\rm i}})^{-\frac{3}{2}\beta_{1}}}
{1+(\frac{t}{t_{\rm i}})^{\frac{3}{4}(1+q)(\beta_{2}-\beta_{1})}}
\end{equation}

\noindent
which is a generalisation of the expression of Wei and Lu (2002) who use the evolution of a regular
cooling break for which $q$ = -1/3. $\beta_1$ and $\beta_2$ are the radiation spectral indices
below and above the injection break respectively.

\begin{figure}
   \centering
\includegraphics[width=8.0cm,height=8.0cm]{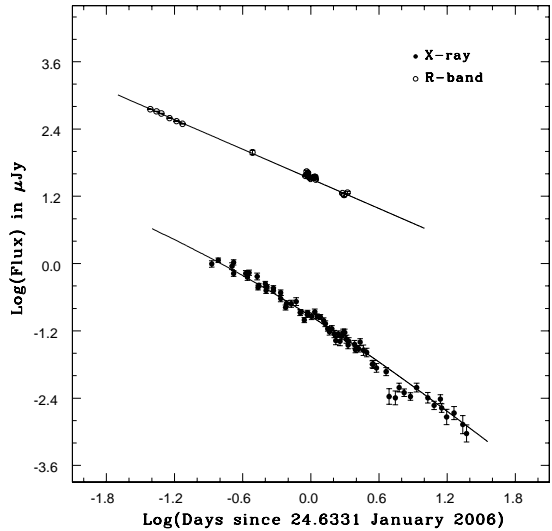}
      \caption{The R band and X-ray light curve of GRB 060124 afterglow. The filled circles represent the X-ray
flux and the open circles the R band flux. The solid line shows the best fit assuming the model described
by Wei and Lu (2002).}
         \label{curved_spectra}
   \end{figure}

\begin{figure}
   \centering
\includegraphics[width=8.0cm,height=8.0cm]{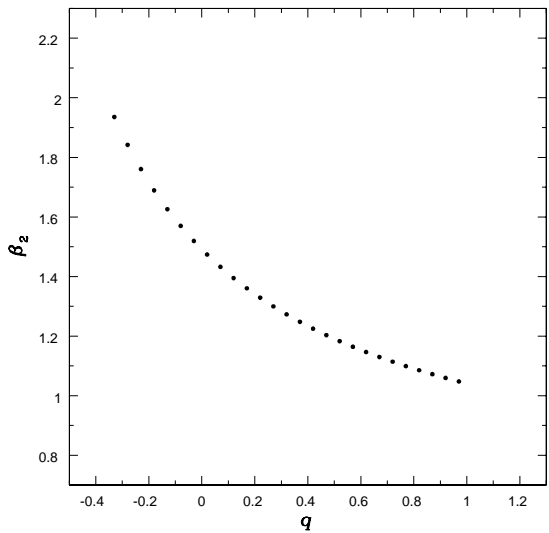}
      \caption{The value of the derived radiation spectral index $\beta_2$ above the injection
break is plotted against $q$, the assumed power law index of dependence of the break energy
on shock Lorentz factor.}
         \label{q_beta2}
   \end{figure}

Based on this model, we have fitted the X-ray and optical afterglow light curve of GRB 060124.
For an assumed $q$ = -1/3, the best fit values obtained for the X-ray light curve are
$t_i$ = 0.55 $\pm$ 0.11 d, $\beta_1$ = 0.59 $\pm$ 0.07 and $\beta_2$ = 1.95 $\pm$ 0.70 with
$\chi^2$/d.o.f. = 2.1. Fits to the R band light curve with the same values of $t_i$, $\beta_1$ and
$\beta_2$ is ruled out at a high confidence level
with a $\chi^2$ / d.o.f. = 225. Thus, the break which is seen in the X-ray light curve is not an achromatic
break.
A good fit with the same $\beta_1$ and $\beta_2$ is obtained for the R band optical light
curve with $\chi^2$/d.o.f. = 3.8 if the optical band break time is
assumed to be $t_i >>$ 2 day, resulting in a single power law evolution with slope $\alpha$ = 3$\beta_1$/2.
The fits to the X-ray and R band light curves are shown in figure \ref{curved_spectra}.
The value of $\alpha$ = 0.89 $\pm$ 0.11 obtained this way is consistent with the single power law fit
to the light curve of all optical bands presented above.
We conclude that the break
in the X-ray band at $\sim$ 10$^5$ sec mentioned by Romano et al. (2006) is not a jet break,
but is possibly due to the steepening of the electron energy spectrum at high energies. The derived value
of the high energy radiation spectral index $\beta_2$ depends on the value of $q$ for which
no independent measurement exists in this case since the passage of the injection break through
multiple frequency bands has not been monitored. The dependence of derived $\beta_2$ on the assumed
value of $q$ in the range -1/3 to 1 is displayed in figure \ref{q_beta2}.

%-----------------------------Figure 4------------------------------%
%_________________________________________________________________________________________

%----------------------------Figure 5-------------------------------%
%_________________________________________________________________________________________

The existence of an injection break within the observed frequency bands have
been inferred for several other GRB afterglows too (see, for example,
Panaitescu and Kumar 2002). This provides an important clue to the
particle acceleration process operating in these afterglows. The
steepening of the injected electron energy spectrum indicates an
upper cutoff, or reduction in efficiency, of the particle acceleration,
which may be due either to radiation losses within the acceleration cycle
time or simply the limited residence time of electrons within the acceleration
zone (Gallant and Achterberg 1999; Gallant et al. 1999).
The appearance of the injection break in the X-ray band around $\sim 1$~day
after the burst implies that the upper cutoff Lorentz factor in this case
lies in the range $10^6$--$10^8$, depending on the fraction of the
total energy in the magnetic field and other physical parameters of the
afterglow. In the present case, good constraints can unfortunately not
be obtained on the physical parameters, in the absence of extensive
multiwavelength coverage. Better constraints on the acceleration
process will be possible if routine, sensitive monitoring of afterglows
is performed at high energies including gamma rays, supported by
extensive long wavelength coverage. It is likely that the
soon-to-be-launched GLAST mission will be able to make significant
impact in this area, particularly for bright afterglows.

The above description of the X-ray and optical light curves implies that within the period of
available observations, i.e. till $\sim$ 2 days after the burst, no jet break has occurred in this
afterglow. While many afterglows exhibit jet break relatively early, those with jet break beyond
$\sim$ 2 days are not rare. Examples include well known afterglows of GRB 000301C, GRB 011211 and
GRB 021004. In fact in a sample of 59 GRB afterglows examined by Zeh et al. (2006), five had jet break
around $\sim$ 2 days and six had jet break between 3 to 8 days after the burst.

The lower limit to the jet break time mentioned above can be used to put constraints on the jet
opening angle and total energy content in the afterglow. Using the isotropic equivalent energy
of 8.9 $\times$ 10$^{52}$ erg derived from $\gamma$ - ray fluence by Cenko et al. (2006) and a lower
limit to the jet break time of 2 days, we estimate the lower limit to the jet opening angle to
be $\sim$ 4.75 degree for an assumed ambient density of 1 atom cm$^{-3}$ and $\gamma$ - ray efficiency
of 0.2. This, in turn, yields a lower limit to the total energy output of 0.31 $\times$ 10$^{51}$ erg,
close to the estimated mean energy output in GRBs (Frail et al. 2001).

It is interesting to note that using the empirical relation (Ghirlanda et al. 2005, Liang \& Zhang 2005)
between isotropic equivalent photon energy output in the burst, the peak energy in the burst and the
jet break time, Romano et al. (2006) predicted a jet break time of (2.1 $\pm$ 1.2) $\times$ 10$^5$ s for
this afterglow, which is consistent with the lower limit to the jet break time discussed by us.

The above discussion demonstrates that observations of GRB afterglows continue to unveil surprising
features in their evolution and reveal aspects of underlying physics that remain to be understood.

\begin{acknowledgements}
We thank the HCT and ARIES observers for kindly sparing their observing time for these observations.
We acknowledge K. Viirone (IAC), J. A. Caballero (IAC, ING) and L. Sabin (ING) for
obtaining the INT image on 15/16 February 2006.
This research has made use of data obtained through the High Energy Astrophysics Science
Archive Research Center Online Service, provided by the NASA/Goddard Space Flight Center.
The GCN system, managed and operated by Scott Barthelmy, is gratefully acknowledged.
One of the authors (KM) thanks L. Resmi and Brijesh Kumar for several useful discussions
while drafting the manuscript. We thank the anonymous referee for constructive comments which
improved the manuscript.
\end{acknowledgements}

%-------------------------------Bibliography--------------------------------%

\end{document}